\documentclass[11pt]{article}

\usepackage{url}            % simple URL typesetting
\usepackage{authblk}
\usepackage{amsfonts}       % blackboard math symbols
\usepackage{graphicx}
\usepackage{amsmath}
\usepackage{amsthm}
\usepackage[ruled]{algorithm2e}
\usepackage{algorithmic}
\usepackage{subfig}
\usepackage{fullpage}

\newtheorem{lemma}{Lemma}

\newtheorem*{theorem*}{Theorem}

\theoremstyle{definition}

\renewcommand{\c}{{\bf c}}

\newcommand{\h}{{\bf h}}
\newcommand{\n}{{\bf n}}

\newcommand{\s}{{\bf s}}

\newcommand{\x}{{\bf x}}
\newcommand{\y}{{\bf y}}
\newcommand{\z}{{\bf z}}

\newcommand{\A}{{\bf A}}

\newcommand{\C}{{\bf C}}

\newcommand{\F}{{\bf F}}

\renewcommand{\H}{{\bf H}}
\newcommand{\I}{{\bf I}}

\newcommand{\N}{{\bf N}}
\renewcommand{\P}{{\bf P}}
\newcommand{\Q}{{\bf Q}}
\newcommand{\R}{\mathbb{R}}
\renewcommand{\S}{{\bf S}}
\newcommand{\U}{{\bf U}}

\newcommand{\W}{{\bf W}}
\newcommand{\X}{{\bf X}}
\newcommand{\Y}{{\bf Y}}
\newcommand{\Z}{{\bf Z}}

\newcommand{\RR}{\mathbb{R}}

\newcommand{\calS}{\mathcal{S}}

\newcommand{\Lam}{\boldsymbol{\Lambda}}

\newcommand{\ol}{\overline}

\DeclareMathOperator{\tr}{Tr}

\newcommand{\argmin}[1]{\underset{#1}{\operatorname{arg}\,\operatorname{min}}\;}

\title{Biologically plausible single-layer networks for nonnegative independent component analysis}

\author[1]{David Lipshutz}
\author[2]{Cengiz Pehlevan}
\author[1,3]{Dmitri B.\ Chklovskii}
\affil[1]{Center for Computational Neuroscience, Flatiron Institute}
\affil[2]{John A. Paulson School of Engineering and Applied Sciences, Harvard University}
\affil[3]{Neuroscience Institute, NYU Medical Center}

\begin{document}

\maketitle

\begin{abstract}
An important problem in neuroscience is to understand how brains extract relevant signals from mixtures of unknown sources, i.e., perform blind source separation. To model how the brain performs this task, we seek a biologically plausible single-layer neural network implementation of a blind source separation algorithm. For biological plausibility, we require the network to satisfy the following three basic properties of neuronal circuits: (i) the network operates in the online setting; (ii) synaptic learning rules are local; (iii) neuronal outputs are nonnegative. Closest is the work by Pehlevan et al.~[Neural Computation, 29, 2925--2954 (2017)], which considers Nonnegative Independent Component Analysis (NICA), a special case of blind source separation that assumes the mixture is a linear combination of uncorrelated, nonnegative sources. They derive an algorithm with a biologically plausible 2-layer network implementation. In this work, we improve upon their result by deriving 2 algorithms for NICA, each with a biologically plausible \textit{single-layer} network implementation. The first algorithm maps onto a network with indirect lateral connections mediated by interneurons. The second algorithm maps onto a network with direct lateral connections and multi-compartmental output neurons. 
\end{abstract}

\section{Introduction}

Brains effortlessly extract relevant signals from mixtures of unknown sources \cite{cherry1953some,desimone1995neural,hulse1997auditory,wilson2006early,narayan2007cortical,bee2008cocktail,shinn2008object,mcdermott2009cocktail,bronkhorst2015cocktail}, an unsupervised signal processing problem known as blind source separation. A classic example in audition is the cocktail party problem, in which a listener tries to follow a single conversation in the presence of multiple background conversations. We seek a model of how brains perform blind source separation.

A special case of blind source separation is Nonnegative Independent Component Analysis (NICA), which assumes a generative model in which the mixture of stimuli is a linear combination of uncorrelated, nonnegative sources; i.e., $\x=\A\s$, where $\s$ denotes the nonnegative vector of source intensities, $\A$ is a mixing matrix and $\x$ denotes the vector of mixed stimuli. The goal of NICA is to infer the source vectors $\s$ from the mixture vectors $\x$. Both the linear additivity of stimuli and nonnegativity of the sources are reasonable assumptions in biological applications. For example, in olfaction, concentrations of odorants are both additive and nonnegative.

\begin{figure}
    \centering
    \includegraphics[width=\textwidth]{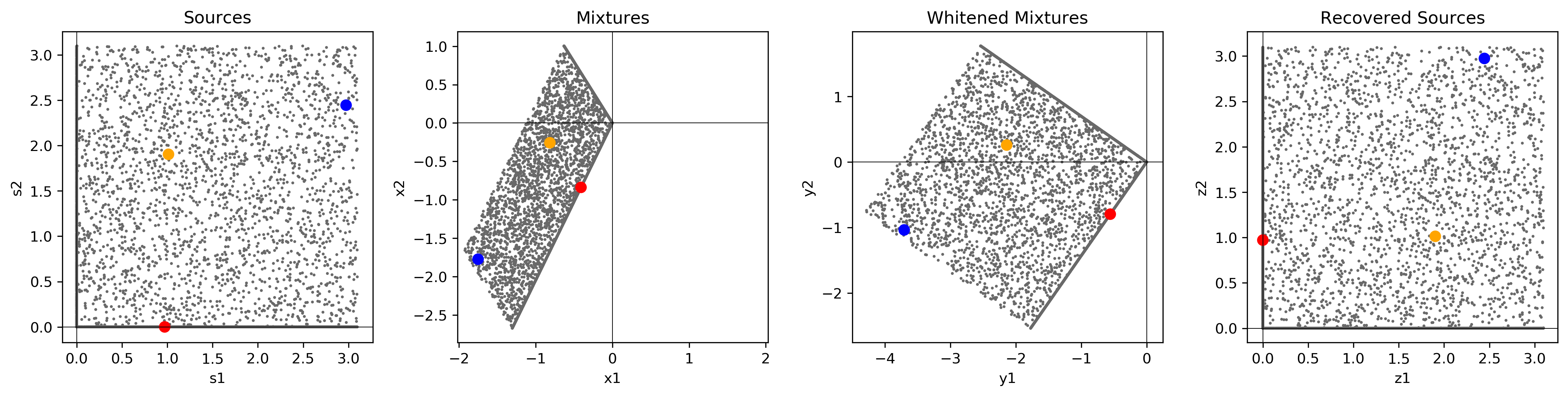}
    \caption{Illustration of Plumbley's 2-step algorithm for NICA. The red, blue and oranges dots track three source vectors across the mixing, whitening and rotation steps. Our algorithms transform Mixtures into Recovered Sources in a single step implemented by single-layer neural networks.}
    \label{fig:scatter}
\end{figure}

Plumbley \cite{plumbley2002conditions} showed that when the sources are well-grounded (i.e., they have nonzero probability of taking infinitesimally small values), NICA can be solved in 2 steps; see Figure~\ref{fig:scatter}. In the first step, the mixture undergoes noncentered whitening; that is, the mixture is linearly transformed to have identity covariance matrix. The second step rotates the mixture until it lies in the nonnegative orthant. The result of these 2 steps must be a permutation of the original sources. This important observation led to a number of algorithms for implementing the rotation step \cite{plumbley2003algorithms,plumbley2004nonnegative,oja2004blind,yuan2004fastica}, many of which have neural network implementations.

Unfortunately, the above-mentioned networks do not offer a viable model of brain function because they do not satisfy one or more of the following three common requirements for biological plausibility \cite{pehlevan2019neuroscience}.
First, the network operates in the online or streaming setting where it receives one input at a time and the output is computed before the next input arrives.
Second, each synaptic update is local in the sense that it depends only on variables represented in the pre- and post-synaptic neurons.
Third, the neuronal outputs are nonnegative.

Building on Plumbley's method, Pehlevan et al.~\cite{pehlevan2017blind} proposed a 2-layer network for NICA, with each layer derived from a principled objective function.
The first layer implements noncentered whitening and the second orthogonally rotates the whitened mixture. While their networks satisfies the requirements for biological plausibility, from a biological perspective, there are advantages to a single-layer network that economizes the number of neurons, which take up valuable resources such as space \cite{rivera2014wiring} and metabolic energy \cite{laughlin2003communication}.

In this work, we derive 2 NICA algorithms (Algorithms~\ref{alg:online1} \& \ref{alg:online2}) that can be implemented in biologically plausible single-layer networks.
The first algorithm maps onto a network with point neurons and indirect lateral connections mediated by interneurons (Figure~\ref{fig:indirect}), and the second algorithm maps onto a network with 2-compartmental neurons and direct lateral connections (Figure~\ref{fig:direct}).
To derive our algorithms, we adopt a normative approach which relies on the fact that the original sources can be expressed (up to permutation) as optimal solutions of single objective functions that combine the 2 objectives from \cite{pehlevan2017blind}.

\paragraph{Notation.} For integers $p,q$, let $\RR^p$ denote $p$-dimensional Euclidean space, $\RR_+^p$ denote the nonnegative orthant in $\RR^p$, $\RR^{p\times q}$ denote the set of $p\times q$ real-valued matrices and $\RR_+^{p\times q}$ denote the subset of matrices with nonnegative entries. 
Let $\calS_{++}^p$ denote the set of $p\times p$ positive definite matrices and let $\I_p$ denote the $p\times p$ identity matrix. Given $T$ samples $\h_1,\dots,\h_T$ of a time series, let
\begin{align*}
    \langle\h\rangle:=\frac1T\sum_{t=1}^T\h_t,&&\C_{hh}:=\frac1T\sum_{t=1}^T(\h_t-\langle\h\rangle)(\h_t-\langle\h\rangle)^\top
\end{align*}
respectively denote the empirical mean and covariance of the time series. Let $\ol\h_t:=\frac1t(\h_1+\cdots+\h_t)$ denote the running sample mean. Given a data matrix $\H=[\h_1,\dots,\h_T]$, let $\delta\H:=[\h_1-\langle\h\rangle,\dots,\h_T-\langle\h\rangle]$ denote the centered data matrix.

\section{Review of prior work}
\label{sec:review}

In this section, we review Plumbley's analysis \cite{plumbley2002conditions} and the objective functions used by Pehlevan et al.~\cite{pehlevan2017blind} to derive a 2-layer network for NICA. Let $d\ge2$ and $\s_1,\dots,\s_T\in\RR_+^d$ be $T$ samples of $d$-dimensional nonnegative source vectors whose components are uncorrelated. Since a constant factor multiplying a source can be absorbed into the associated column of the mixing matrix $\A$, we can assume, without loss of generality, that each component of the source vector has unit sample variance. In particular, $\C_{ss}=\I_d$. Let $k\ge d$, $\A$ be a full rank $k\times d$ mixing matrix and define the $k$-dimensional mixture vectors by $\x_t:=\A\s_t$ for $t=1,\dots,T$.

\subsection{Plumbley's NICA method}

Plumbley \cite{plumbley2002conditions} proposed solving NICA in 2 steps: noncentered whitening followed by orthogonal transformation, which are depicted in Figure~\ref{fig:scatter}.

Noncentered whitening is a linear transformation $\y:=\F\x$ of the mixture, where $\y\in\R^d$ and $\F$ is a $d \times k$ whitening matrix such that $\y$ has identity covariance matrix, i.e., $\C_{yy}=\I_d$. The combined effect of source mixing and prewhitening steps, which is encoded in the $d\times d$ matrix $\F\A$ (since $\y=\F\x$ and $\x=\A\s$), is an orthogonal transformation. 
To see this, we use the facts that $\C_{ss}=\I_d$, $\y=\F\A\s$ and $\C_{yy}=\I_d$ to write 
\begin{align*}
    (\F\A)(\F\A)^\top=(\F\A)\C_{ss}(\F\A)^\top&=\frac1T\sum_{t=1}^T\F\A(\s_t-\langle\s\rangle)(\s_t-\langle\s\rangle)^\top(\F\A)^\top\\
    &=\frac1T\sum_{t=1}^T(\y_t-\langle\y\rangle)(\y_t-\langle\y\rangle)^\top=\C_{yy}=\I_d.
\end{align*}

In the second step, one looks for an orthogonal matrix ${\bf R}$ such that the transformation $\z:={\bf R}\y$ is nonnegative. 
For the solution to be unique up to a permutation, each source $s^i$ must be well grounded; that is, $P(s^i<\delta)>0$ for all $\delta>0$. 
Then by \cite[Theorem 1]{plumbley2002conditions}, the vector $\z$ is equal to a permutation of the sources $\s$.

\subsection{Similarity matching objectives for the 2-step algorithm}

To obtain a biologically plausible network, Pehlevan et al.~\cite{pehlevan2017blind} proposed novel mathematical formulations of the noncentered whitening and rotation steps, which can be  implemented by a biological plausible 2-layer network.

Here we recall the principled objective functions they use for each layer, which are closely related to the objective functions we use to derive our networks.
To this end, define the $k\times T$ concatenated data matrix $\X:=[\x_1,\dots,\x_T]$.
In the first step, Pehlevan et al.~\cite{pehlevan2017blind} optimize, with respect to the $d\times T$ matrix $\Y:=[\y_1,\dots,\y_T]$, the following objective:
\begin{align}\label{eq:smXH}
    \argmin{\Y\in\R^{d\times T}}-\tr(\delta\Y^\top\delta\Y \delta\X^\top\delta\X)\quad\text{subject to}\quad\delta\Y^\top \delta\Y\preceq T\I_T\quad\text{and}\quad\Y=\F\X,
\end{align}
where $\F$ is some $d\times k$ matrix and the constraint enforces that the difference $T\I_T-\delta\Y^\top \delta\Y$ is positive semidefinite.  
As shown in \cite{pehlevan2017blind}, objective \eqref{eq:smXH} is optimized when $\Y$ is a noncentered whitened transformation of $\X$.

For the second step, Pehlevan et al.~\cite{pehlevan2017blind} introduce the following Nonnegative Similarity Matching (NSM) objective:
\begin{align}\label{eq:nsmH}
    \argmin{\Z\in\R_+^{d\times T}}\|\Z^\top\Z-\Y^\top\Y\|_{\text{Frob}}^2.
\end{align}
The objective minimizes the mismatch between similarities of the nonnegative outputs $\Z$ and the noncentered whitened mixtures $\Y$ (as measured by inner products).
As shown in \cite{pehlevan2017blind}, any orthogonal transformation of $\Y$ to the nonnegative orthant, which corresponds to a permutation of the original sources, is a solution of the NSM objective \eqref{eq:nsmH}.

From objectives \eqref{eq:smXH} and \eqref{eq:nsmH}, Pehlevan et al.~\cite{pehlevan2017blind} derive a 2-step algorithm for NICA that can be implemented in a 2-layer neural network that operates in the online setting, uses local learning rules, and whose rotation layer has nonnegative neuronal outputs.

\section{Combined objectives for NICA}

We now modify objectives \eqref{eq:smXH} and \eqref{eq:nsmH} to obtain 2 objectives for NICA, which will be the starting points for the derivations of our 2 online NICA algorithms with single-layer neural network implementations.

\subsection{Adding a nonnegativity constraint to the noncentered whitening objective}

We first modify the noncentered whitening objective \eqref{eq:smXH}.
Note that the solution of objective \eqref{eq:smXH} is not unique --- left multiplying any solution $\Y$ by an orthogonal matrix ${\bf R}$ yields another noncentered whitened transformation of $\X$.
In fact, the second step of Plumbley's method \cite{plumbley2002conditions} is to identify an orthogonal transformation ${\bf R}$ that results in a \textit{nonnegative} whitened transformation $\Z={\bf R}\Y$.
Here, we combine the 2 objectives by adding a nonnegativity constraint to the noncentered whitening objective \eqref{eq:smXH}, as follows:
\begin{align}\label{eq:smXHmod}
    \argmin{\Y\in\R_+^{d\times T}}-\tr(\delta\Y^\top\delta\Y\delta\X^\top\delta\X)\quad\text{subject to}\quad\delta\Y^\top\delta\Y\preceq T\I_T\quad\text{and}\quad\Y=\F\X,
\end{align}
where $\F$ is some $d\times k$ matrix.

\subsection{Adding a whitening matrix to the NSM objective}

Next, we alter the NSM objective \eqref{eq:nsmH} by replacing the Gram matrix $\Y^\top\Y$ with terms that depend only on $\X$, which will avoid the need for the noncentered whitening step. Consider the eigendecomposition of the covariance matrix $\C_{xx}=\U\Lam\U^\top$, where $\U$ is a $k\times d$ matrix with orthonormal column vectors and $\Lam$ is a $d\times d$ diagonal matrix whose diagonal entries are the nonzero eigenvalues of $\C_{xx}$. Then the whitening matrix $\F$ must be of the form $\Q\Lam^{-\frac12}\U^\top$, where $\Q$ is an arbitrary $d\times d$ orthogonal matrix. Therefore,
\begin{align*}
    \Y^\top\Y=\X^\top\F^\top\F\X=\X^\top\U\Lam^{-1}\U^\top\X=\X^\top\C_{xx}^+\X,
\end{align*}
where $\C_{xx}^+:=\U\Lam^{-1}\U^\top$ is the Moore-Penrose inverse of $\C_{xx}$. Substituting in for $\Y^\top\Y$ in the NSM objective \eqref{eq:nsmH} results in our second objective:
\begin{align}
\label{eq:nsm}
    \argmin{\Z\in\R_+^{d\times T}}\|\Z^\top\Z-\X^\top\C_{xx}^+\X\|_{\text{Frob}}^2.
\end{align}

\section{Single-layer neural networks for NICA}
\label{sec:NICA}

Starting from objectives \eqref{eq:smXHmod} and \eqref{eq:nsm}, we derive our 2 online NICA algorithms.
The first algorithm maps onto a single-layer network with point neurons and \textit{indirect} lateral connections.
The second algorithm maps onto a single-layer network with 2-compartmental neurons and \textit{direct} lateral connections.

\subsection{Single-layer network with point neurons and indirect lateral connections}
\label{sec:NICA1}

The derivation of our online algorithm starting from objective \eqref{eq:smXHmod} closely follows the derivation of the whitening layer in the network derived in \cite{pehlevan2017blind}. The main difference is that the neuronal outputs are constrained to be nonnegative. To begin, we introduce $m$-dimensional activity vectors $\n_1,\dots,\n_T$, with $m\ge d$, which we concatenate into the data matrix $\N:=[\n_1,\dots,\n_T]$, and use the Gramian $\delta\N^\top\delta\N$ as a Lagrange multiplier to enforce the constraint $\delta\Y^\top\delta\Y\preceq T\I_T$:
\begin{align*}
    \min_{\Y\in\R_+^{d\times T}}\max_{\N\in\R^{m
    \times T}}\tr\left[-\delta\Y^\top\delta\Y\delta\X^\top\delta\X+\delta\N^\top\delta\N(\delta\Y^\top\delta\Y-T\I_T)\right]\quad\text{subject to}\quad\Y=\F\X.
\end{align*}
Next, we normalize by $T^2$ and substitute synaptic weight matrices $\W_{xy}$ and $\W_{yn}$ in place of $\frac1T\delta\Y\delta\X^\top$ and $\frac1T\delta\N\delta\Y^\top$, respectively:
\begin{align*}
    \min_{\Y\in\R_+^{d\times T}}\max_{\N\in\RR^{d
    \times T}}\min_{\W_{xy}\in\RR^{d\times k}}\max_{\W_{yn}\in\RR^{m\times d}}L_1(\Y,\N,\W_{xy},\W_{yn})\quad\text{subject to}\quad\Y=\F\X,
\end{align*}
where
\begin{align*}
    L_1(\Y,\N,\W_{xy},\W_{yn})&:=\frac1T\tr\left(2\delta\N^\top\W_{yn}\delta\Y-2\delta\Y^\top\W_{xy}\delta\X-\delta\N^\top\delta\N\right)\\
    &\qquad-\tr\left(\W_{yn}\W_{yn}^\top)+\tr(\W_{xy}\W_{xy}^\top\right).
\end{align*}
The substitution can be readily justified by differentiating $L_1$ with respect to $\W_{xy}$ and $\W_{yn}$ and noting the minimum (resp.\ maximum) is achieved when $\W_{xy}=\tfrac1T\delta\Y\delta\X^\top$ (resp.\ $\W_{yn}=\tfrac1T\delta\N\delta\Y^\top)$. Since $L_1$ satisfies the saddle point property with respect to $\N$ and $\W_{xy}$, and with respect to $\Y$ and $\W_{yn}$, we can interchange the order of optimization, as follows:
\begin{align}\label{eq:minmax2}
    \min_{\W_{xy}\in\RR^{d\times k}}\max_{\W_{yn}\in\RR^{m\times d}}\min_{\Y\in\R_+^{d\times T}}\max_{\N\in\RR^{d
    \times T}}L_1(\Y,\N,\W_{xy},\W_{yn})\quad\text{subject to}\quad\Y=\F\X.
\end{align}
We first solve objective \eqref{eq:minmax2} in the offline setting. In general, optimizing over $(\Y,\N)$ is challenging due to the constraint that $\Y$ be a nonnegative linear transformation of $\X$. In appendix \ref{apdx:derivation}, we show that when the synaptic weights $\W_{xy}$ and $\W_{yn}$ are at their optimal values, we can optimize over $(\Y,\N)$ by repeating the following projected gradient descent steps until convergence:
\begin{align}\label{eq:Yoffline}
    \Y&\gets\left[\Y+\gamma\left(\W_{xy}\X-\W_{yn}^\top\N\right)\right]_+,&&\N\gets\N+\gamma\left(\W_{yn}\Y-\N\right),
\end{align}
where $\gamma>0$ is a small step size and $[\cdot]_+$ denotes taking the positive part elementwise, which ensures the nonnegativity of $\Y$.
In the case the synaptic weights $\W_{xy}$ and $\W_{yn}$ are not at their optimal values, we repeat the above projected gradient descent steps until convergence to obtain an approximation of the optimal $(\Y,\N)$.
We then perform a gradient descent-ascent step of the objective $L_1$ with respect to $\W_{xy}$ and $\W_{yn}$:
\begin{align}
    \label{eq:dWxy}
    \W_{xy}&\gets\W_{xy}+\eta\left(\frac1T\delta\Y\delta\X^\top-\W_{xy}\right)\\
    \label{eq:dWyn}
    \W_{yn}&\gets\W_{yn}+\eta\left(\frac1T\delta\N\delta\Y^\top-\W_{yn}\right).
\end{align}
Here $\eta>0$ is the learning rate for $\W_{xy}$ and $\W_{yn}$.

Next, we solve the objective \eqref{eq:minmax2} in the online setting.
At each time step $t$, we approximate the optimization over $(\y_t,\n_t)$ by taking the following projected gradient descent steps until convergence:
\begin{align}\label{eq:ynupdate}
    \y_t&\gets[\y_t+\gamma(\W_{xy}\x_t-\W_{ny}\n_t)]_+,&&\n_t\gets\n_t+\gamma(\W_{yn}\y_t-\n_t),
\end{align}
where we have defined $\W_{ny}:=\W_{yn}^\top$. We then take stochastic gradient descent-ascent steps in $\W_{xy}$ and $\W_{yn}$ by replacing the averages in equations~\eqref{eq:dWxy} and \eqref{eq:dWyn} with their online approximations:
\begin{align*}
    \W_{xy}&\gets\W_{xy}+\eta\left((\y_t-\ol\y_t)(\x_t-\ol\x_t)^\top-\W_{xy}\right)\\
    \W_{yn}&\gets\W_{yn}+\eta\left((\n_t-\ol\n_t)(\y_t-\ol\y_t)^\top-\W_{yn}\right)\\
    \W_{ny}&\gets\W_{ny}+\eta\left((\y_t-\ol\y_t)(\n_t-\ol\n_t)^\top-\W_{ny}\right).
\end{align*}
The symmetry of the updates for $\W_{ny}$ and $\W_{yn}$ ensures that $\W_{ny}=\W_{yn}^\top$ after each iteration provided the constraint holds at initialization.
As we show in appendix~\ref{sec:weight}, we can relax this initialization constraint, which yields our first online NICA algorithm, Algorithm~\ref{alg:online1}.

\vspace{.5cm}
\begin{algorithm}[H]
\caption{Bio-NICA with interneurons}
\label{alg:online1}
\begin{algorithmic}
    \STATE {\bfseries input} mixtures $\{\x_1,\dots,\x_T\}$; parameters $\gamma$, $\eta$
    \STATE {\bfseries initialize} $\W_{xy}$, $\W_{yn}$, $\W_{ny}$, $\ol\x_0={\bf 0}$, $\ol\y_0={\bf 0}$, $\ol\n_0={\bf 0}$
    \FOR{$t=1,2,\dots,T$}
        \REPEAT 
            \STATE $\y_t\gets [\y_t+\gamma(\W_{xy}\x_t-\W_{ny}\n_t)]_+$
            \STATE $\n_t\gets \n_t+\gamma(\W_{yn}\y_t-\n_t)$
        \UNTIL{convergence}
        \STATE $\ol\x_t\gets\ol\x_{t-1}+\frac1t(\x_t-\ol\x_{t-1})$
        \STATE $\ol\y_t\gets\ol\y_{t-1}+\frac1t(\y_t-\ol\y_{t-1})$
        \STATE $\ol\n_t\gets\ol\n_{t-1}+\frac1t(\n_t-\ol\n_{t-1})$
        \STATE $\W_{xy} \gets\W_{xy}+\eta((\y_t-\ol\y_t)(\x_t-\ol\x_t)^\top-\W_{xy})$
        \STATE $\W_{ny} \gets\W_{ny}+\eta((\y_t-\ol\y_t)(\n_t-\ol\n_t)^\top-\W_{ny})$
        \STATE $\W_{yn} \gets\W_{yn}+\eta((\n_t-\ol\n_t)(\y_t-\ol\y_t)^\top-\W_{yn})$
    \ENDFOR
\end{algorithmic}
\end{algorithm}
\vspace{.5cm}

\begin{figure}
    \centering
    \includegraphics[width=.5\textwidth]{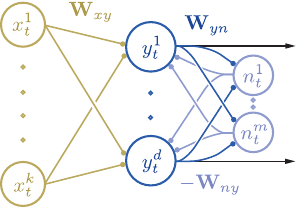}
    \caption{Single-layer network with interneurons for implementing Algorithm~\ref{alg:online1}.}
    \label{fig:indirect}
\end{figure}

Algorithm \ref{alg:online1} can be implemented in a single-layer network with point neurons and indirect lateral connections mediated by interneurons, Figure~\ref{fig:indirect}, so we refer to the algorithm as `Bio-NICA with interneurons'. The network consists of $k$ input neurons, $d$ principal (output) output neurons and $m$ interneurons. Feedforward synapses between the input and principal neurons encode the weight matrix $\W_{xy}$ and lateral synapses between the principal neurons (resp.\ interneurons) and the interneurons (resp.\ principal neurons) encode the weight matrix $\W_{yn}$ (resp.\ $\W_{ny}$). At each time step $t$, the $k$-dimensional mixture $\x_t$, which is represented by the $k$ input neurons, is multiplied by the weight matrix $\W_{xy}$, which yields the $d$-dimensional projection $\W_{xy}\x_t$. This is followed by the fast recurrent dynamics in equation~\eqref{eq:ynupdate}. The equilibrium values of $\y_t$ and $\n_t$ respectively correspond to the nonnegative output of the principal neurons and the output of the interneurons.

We can write the elementwise synaptic updates as follows,
\begin{align*}
    W_{xy}^{ij}&\gets W_{xy}^{ij}+\eta\left((y_t^i-\ol y_t^i)(x_t^j-\ol x_t^j)-W_{xy}^{ij}\right),&&1\le i\le d,\;1\le j\le k,\\
    W_{ny}^{ij}&\gets W_{ny}^{ij}+\eta\left((y_t^i-\ol y_t^i)(n_t^j-\ol n_t^j)-W_{ny}^{ij}\right),&&1\le i\le d,\;1\le j\le m,\\
    W_{yn}^{ij}&\gets W_{yn}^{ij}+\eta\left((n_t^i-\ol n_t^i)(y_t^j-\ol y_t^j)-W_{yn}^{ij}\right),&&1\le i\le m,\;1\le j\le d,
\end{align*}
where we recall that $\ol\x_t$, $\ol\y_t$ and $\ol\n_t$ are the running means of $\x_t$, $\y_t$ and $\n_t$, respectively. We assume that each neuron stores the running mean of its activity. Biologically, these means could be represented at the pre- and post-synaptic terminals by slowly changing calcium concentrations. From the elementwise updates, we see that the update for each synapse is local in the sense that it only depends on variables that are represented in the pre- and post-synaptic neurons.

\subsection{Single-layer network with 2-compartmental neurons and direct lateral connections}
\label{sec:NICA2}

The derivation of the our online algorithm starting form objective \eqref{eq:nsm} is closely related to the derivation of the single-layer networks with multi-compartmental neurons for solving generalized eigenvalue problems \cite{lipshutz2020biologically,lipshutz2021biologically}. To begin, we expand the square, drop terms that do not depend on $\Z$, and normalize by $T^2$:
\begin{align}\label{eq:lateral_expand}
    \min_{\Z\in\RR_+^{d\times T}}\frac{1}{T^2}\tr\left(-2\Z^\top\Z\X^\top\C_{xx}^+\X+\Z^\top\Z\Z^\top\Z\right).
\end{align}
Next, we introduce synaptic weight matrices $\W_{xz}$ and $\W_{zz}$ in place of $\frac1T\Z\X^\top\C_{xx}^+$ and $\frac1T\Z\Z^\top$, respectively, which results in the minimax objective:
\begin{align}\label{eq:minmax1}
    \min_{\Z\in\RR_+^{d\times T}}\min_{\W_{xz}\in\RR^{d\times k}}\max_{\W_{zz}\in\mathcal{S}_{++}^d}L_2(\Z,\W_{xz},\W_{zz}),
\end{align}
where
\begin{align*}
    L_2(\Z,\W_{xz},\W_{zz}):=\frac2T\tr\left(\Z^\top\W_{zz}\Z-2\Z^\top\W_{xz}\X\right)-\tr\left(\W_{zz}^2-2\W_{xz}\C_{xx}\W_{xz}^\top\right).
\end{align*}
The equivalence between the minimization problem \eqref{eq:lateral_expand} and the minimax problem \eqref{eq:minmax1} can be seen by taking partial derivatives of $L_2$ with respect to $\W_{xz}$ (resp.\ $\W_{zz}$) and noting the minimum (resp.\ maximum) is achieved when $\W_{xz}=\tfrac1T\Z\X^\top\C_{xx}^+$ (resp.\ $\W_{zz}=\tfrac1T\Z\Z^\top)$. Since the objective $L_2$ satisfies the strict saddle point property with respect to $\Z$ and $\W_{zz}$, we can interchange the order of optimization, as follows:
\begin{align}\label{eq:minmax}
    \min_{\W_{xz}\in\RR^{d\times k}}\max_{\W_{zz}\in\mathcal{S}_{++}^d}\min_{\Z\in\RR_+^{d\times T}}L_2(\Z,\W_{xz},\W_{zz}).
\end{align}
We first solve the minimax objective \eqref{eq:minmax} in the offline setting by minimizing $L_2$ over $\Z$ and then taking gradient descent-ascent steps in $\W_{xz}$ and $\W_{zz}$.
The minimization over $\Z$ can be approximated by repeating the following projected gradient descent steps until convergence:
\begin{align*}
    \Z\gets[\Z+\gamma(\W_{xz}\X-\W_{zz}\Z)]_+,
\end{align*}
where $\gamma>0$ is a small step size.
Next, having minimized over $\Z$, we perform a gradient descent-ascent step of the objective function $L_2$ with respect to $\W_{xz}$ and $\W_{zz}$:
\begin{align}
    \label{eq:dWoff}
    \W_{xz}&\gets\W_{xz}+2\eta\left(\frac1T\Z\X^\top-\W_{xz}\C_{xx}\right),\\
    \label{eq:dMoff}
    \W_{zz}&\gets\W_{zz}+\frac{\eta}{\tau}\left(\frac1T\Z\Z^\top-\W_{zz}\right).
\end{align}
Here $\tau>0$ is the ratio between the learning rates for $\W_{xz}$ and $\W_{zz}$, and $\eta\in(0,\tau)$ is the learning rate for $\W_{xz}$.
The upper bound $\eta<\tau$ ensures that $\W_{zz}$ remains positive definite given a positive definite initialization.

To solve the minimax objective \eqref{eq:minmax} in the online setting, we take stochastic gradient ascent-descent steps.
At each time step $t$, analogous to the offline setting, we first minimize over the output $\z_t$ by repeating the following projected gradient descent steps until convergence:
\begin{align}
\label{eq:ztupdate}
    \z_t\gets[\z_t+\gamma(\c_t-\W_{zz}\z_t)]_+,
\end{align}
where we have defined the projection $\c_t:=\W_{xz}\x_t$.
We then take stochastic gradient descent-ascent steps in $\W_{xz}$ and $\W_{zz}$. 
To this end, we replace the averages $\tfrac1T\Z\X^\top$ and $\tfrac1T\Z\Z^\top$ in equations~\eqref{eq:dWoff} and \eqref{eq:dMoff} with their respective online approximations $(\z_t-\ol\z_t)(\x_t-\ol\x_t)^\top$ and $(\z_t-\ol\z_t)(\z_t-\ol\z_t)^\top$. While we could approximate the matrix $\W_{xz}\C_{xx}$ in the online setting with $\W_{xz}(\x_t-\ol\x_t)(\x_t-\ol\x_t)$, this does not lead to local learning rules. Instead, we observe that
\begin{align*}
    \W_{xz}\C_{xx}=\frac1T\sum_{t=1}^T\W_{xz}(\x_t-\langle\x\rangle)(\x_t-\langle\x\rangle)^\top=\frac1T\sum_{t=1}^T(\c_t-\langle\c\rangle)(\x_t-\langle\x\rangle)^\top,
\end{align*}
and replace $\W_{xz}\C_{xx}$ with the online approximation $(\c_t-\ol\c_t)(\x_t-\ol\x_t)^\top$.
This yields our second online algorithm for NICA, Algorithm~\ref{alg:online2}.

\vspace{.5cm}
\begin{algorithm}[H]
\caption{Bio-NICA with 2-compartmental neurons}
\label{alg:online2}
\begin{algorithmic}
    \STATE {\bfseries input} mixtures $\{\x_1,\dots,\x_T\}$; parameters $\gamma$, $\eta$, $\tau$
    \STATE {\bfseries initialize} $\W_{xz}$, $\W_{zz}$, $\ol\x_0={\bf 0}$, $\ol\c_0={\bf 0}$
    \FOR{$t=1,2,\dots,T$}
        \STATE $\c_t\gets\W_{xz}\x_t$
        \REPEAT 
            \STATE $\z_t\gets [\z_t+\gamma(\c_t-\W_{zz}\z_t)]_+$
        \UNTIL{convergence}
        \STATE $\ol\x_t\gets\ol\x_{t-1}+\frac1t(\x_t-\ol\x_{t-1})$
        \STATE $\ol\c_t\gets\ol\c_{t-1}+\frac1t(\c_t-\ol\c_{t-1})$
        \STATE $\W_{xz} \gets\W_{xz}+2\eta(\z_t\x_t^\top-(\c_t-\ol\c_t)(\x_t-\ol\x_t)^\top)$
        \STATE $\W_{zz} \gets\W_{zz}+\frac{\eta}{\tau}(\z_t\z_t^\top-\W_{zz})$
    \ENDFOR
\end{algorithmic}
\end{algorithm}
\vspace{.5cm}

\begin{figure}
\centering
    \includegraphics[width=.5\textwidth]{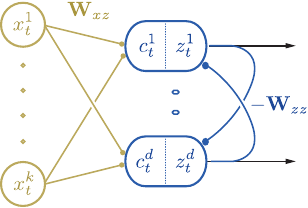}
    \caption{Single-layer network with 2-compartmental neurons for implementing Algorithm~\ref{alg:online2}.}
    \label{fig:direct}
\end{figure}

Algorithm \ref{alg:online2} can be implemented in a single-layer network with 2-compartmental neurons and direct lateral connections, Figure~\ref{fig:direct}, so we refer to the algorithm as `Bio-NICA with 2-compartmental neurons'. The network consists of $k$ input neurons and $d$ output neurons. Each output neuron has a dendritic compartment and a somatic compartment. Feedforward synapses between the input and output neurons encode the weight matrix $\W_{xz}$ and recursive lateral synapses between the output neurons encode the weight matrix $-\W_{zz}$. At each time step $t$, the $k$-dimensional mixture $\x_t$, which is represented by the input neurons, is multiplied by the weight matrix $\W_{xz}$, which is encoded by the feedforward synapses connecting the input neurons to the output neurons. This yields the $d$-dimensional projection $\c_t=\W_{xz}\x_t$, which is computed in the dendritic compartments of the output neurons and then propagated to their somatic compartments. This is followed by the fast recurrent neural dynamics in equation \eqref{eq:ztupdate}. The equilibrium value of $\z_t$ corresponds to the nonnegative somatic activity of the output neurons.

The elementwise synaptic updates are as follows,
\begin{align*}
    W_{xz}^{ij}&\gets W_{xz}^{ij}+2\eta\left(z_t^ix_t^j-(c_t^i-\ol c_t^i)(x_t^j-\ol x_t^j)\right), && 1\le i\le d,\;1\le j\le k,\\
    W_{zz}^{ij}&\gets W_{zz}^{ij}+\frac{\eta}{\tau}\left(z_t^iz_t^j-W_{zz}^{ij}\right), && 1\le i,j\le d,
\end{align*}
where we recall that $\ol\x_t$ and $\ol\c_t$ are the running means of $\x_t$ and $\c_t$, respectively. We assume that the input neurons and output neurons respectively store the running means $\ol\x_t$ and $\ol\c_t$. Thus, we see that the update for each synapse is local; that is, the update depends only on variables that are represented in the pre- and post-synaptic neurons.

\section{Numerical experiments}
\label{sec:numerics}

We evaluated Algorithms~\ref{alg:online1} and \ref{alg:online2} on synthetic and real datasets and compare their performance to 2 state-of-the-art online NICA algorithms: Nonnegative PCA \cite{plumbley2004nonnegative} and 2-layer NSM \cite{pehlevan2017blind}. Nonnegative PCA requires (noncentered) pre-whitened inputs, which we implemented offline. To quantify the performance of the algorithms, we use the mean-squared error,
\begin{align*}
    \text{error}(t)=\frac{1}{td}\sum_{t'=1}^t|\s_{t'}-\P\y_{t'}|^2,
\end{align*}
where $\P$ is the permutation matrix that minimizes the error at the final time point.
For detailed descriptions of our implementations, see appendix~\ref{apdx:numerics}. The evaluation code is available at \url{https://github.com/flatironinstitute/bio-nica}.

\subsection{Mixture of sparse random uniform sources}
\label{sec:uniform}

\begin{figure}
    \centering
    \includegraphics[width=.48\textwidth]{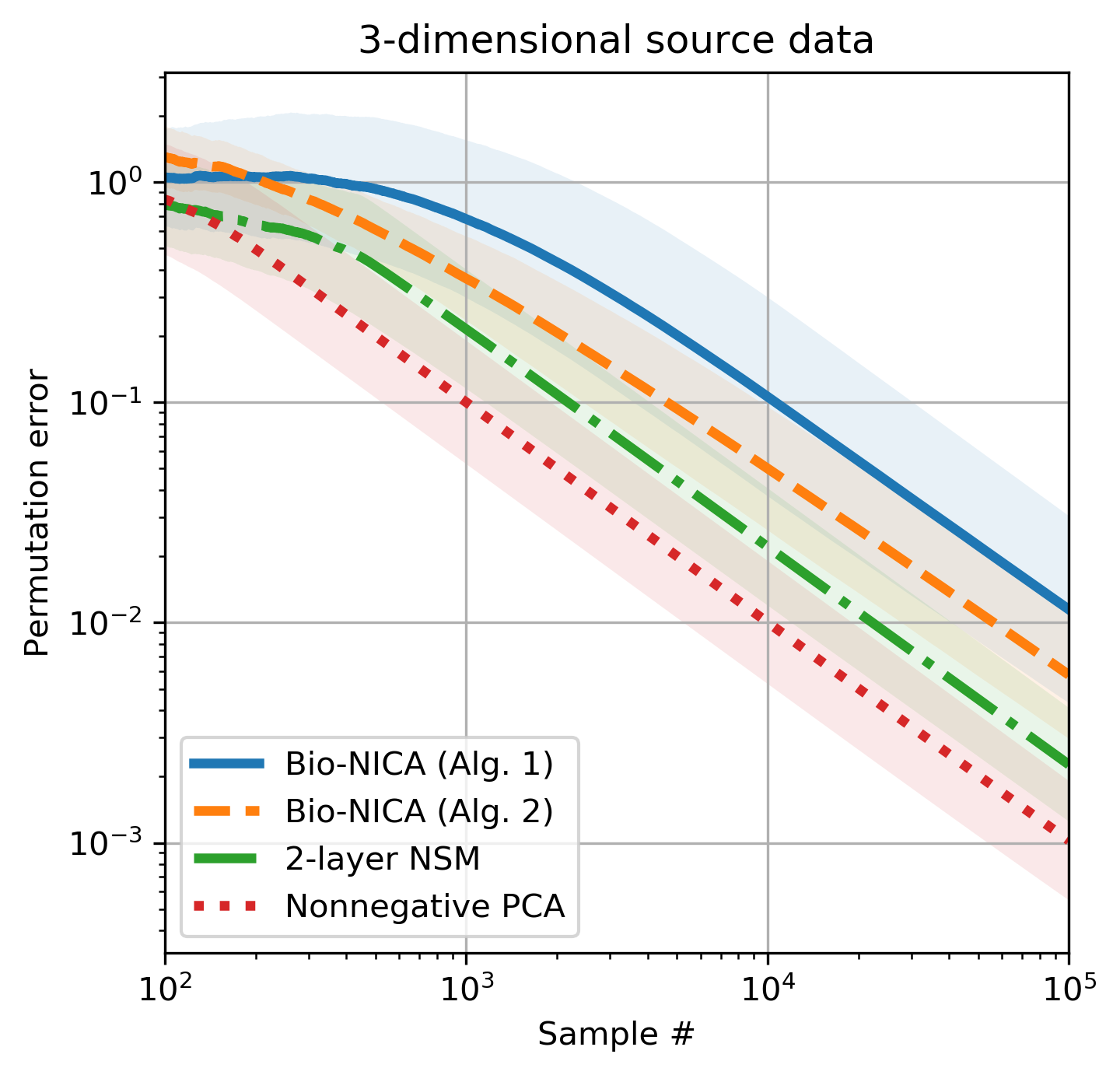}\hfill\includegraphics[width=.48\textwidth]{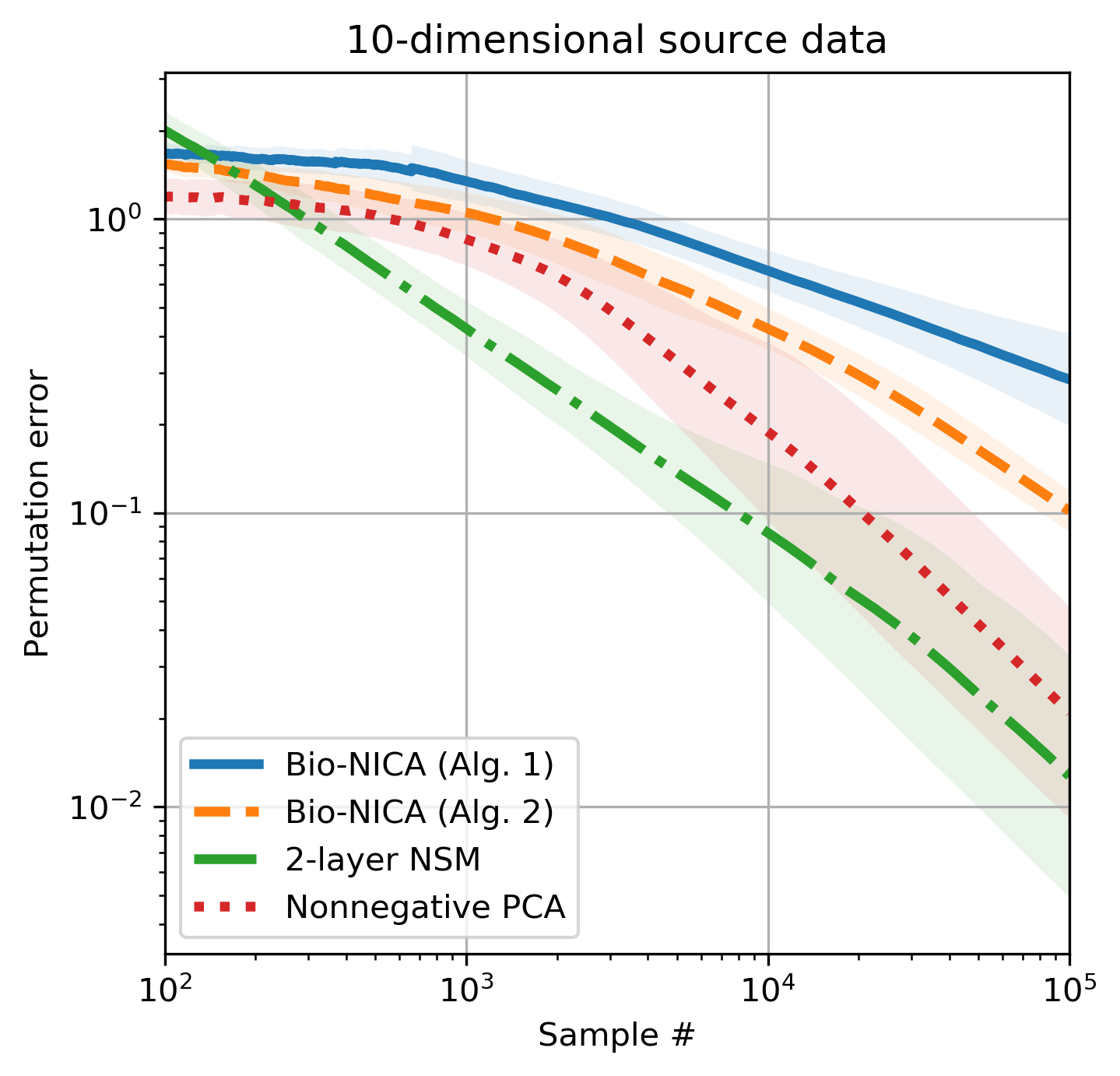}
    \caption{Performance of algorithms when presented with mixtures of sparse random uniform sources, in terms of permutation error. The lines and shaded regions denote the means and 90\% confidence intervals over 10 runs.}
    \label{fig:comparison}
\end{figure}

We first compare the algorithms on a synthetic dataset generated by independent and identically distributed samples.
Following \cite{pehlevan2017blind}, each source sample was set to zero with probability $1/2$ or sampled uniformly from the interval $(0,\sqrt{48/5})$ with probability $1/2$.
We used random square mixing matrices whose elements were independent standard normal random variables.
In Figure~\ref{fig:comparison}, we plot the performance of each algorithm on mixtures of 3- and 10-dimensional sources.

\subsection{Mixture of natural images}

\begin{figure}
    \centering
    \includegraphics[width=.545\textwidth]{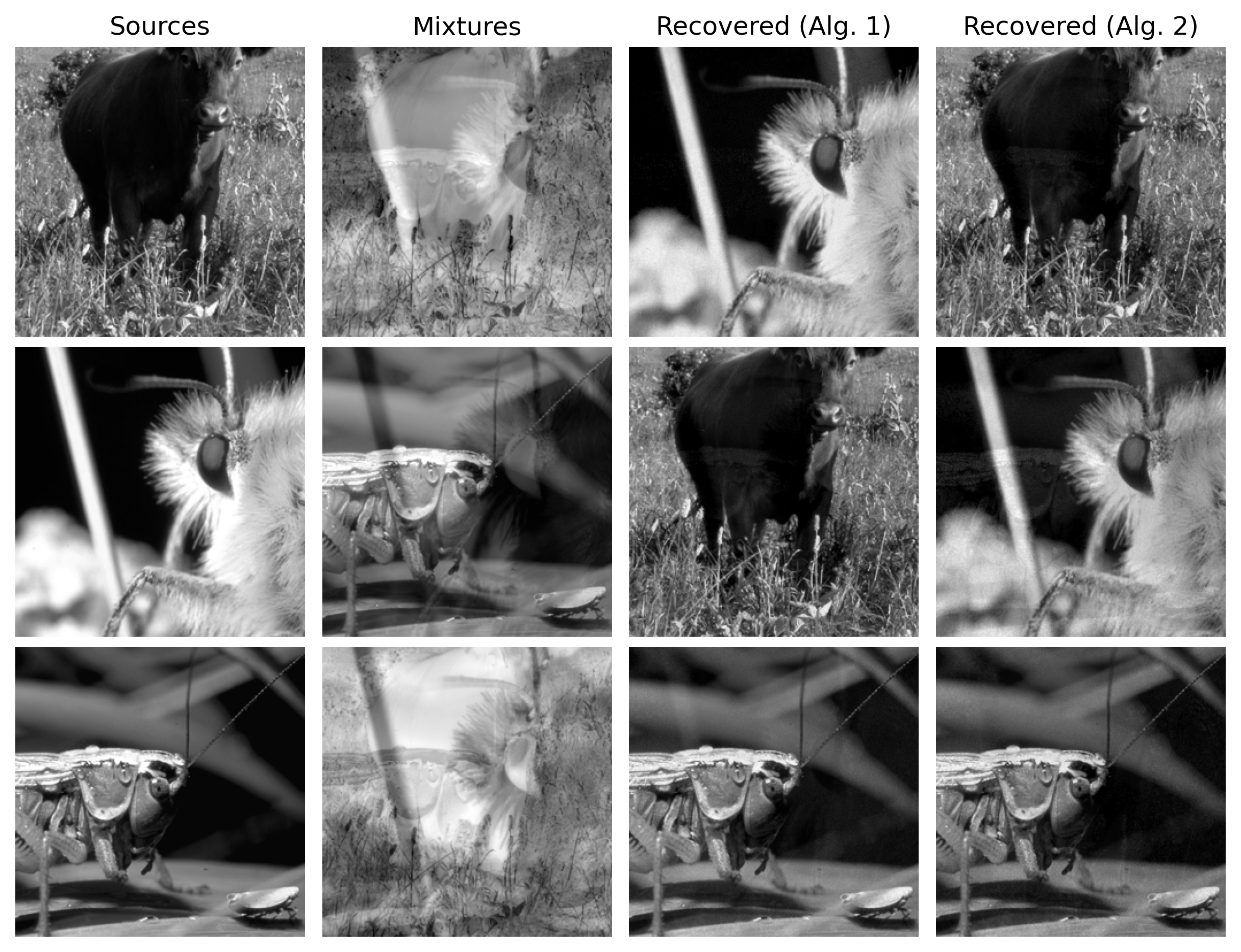}\hfill
    \includegraphics[width=.425\textwidth]{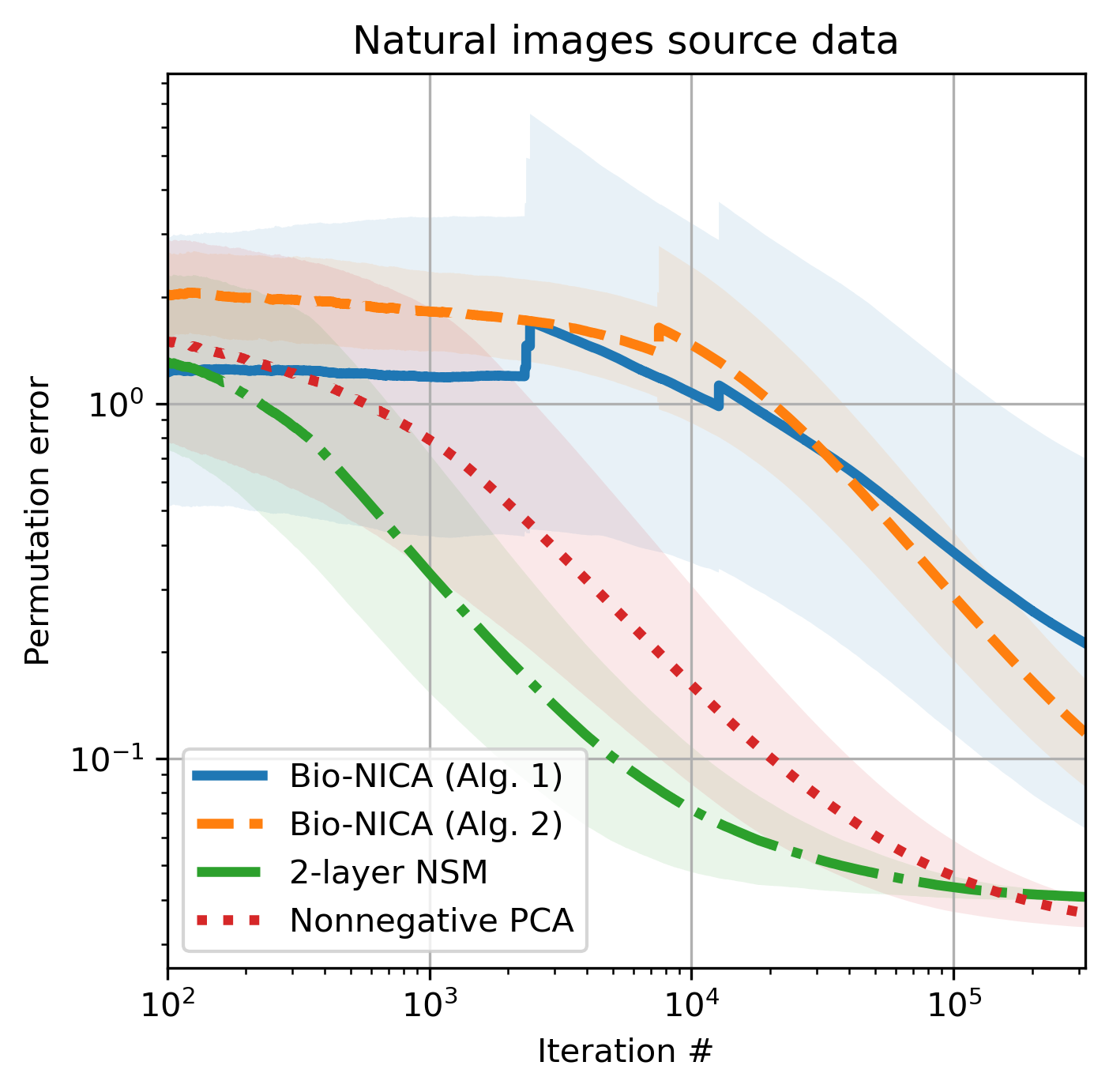}
    \caption{Performance of algorithms when presented with mixtures of natural images. The left image shows the sources, mixtures, and recovered sources (from Algorithms~\ref{alg:online1} and \ref{alg:online2}). The right plot shows the performance of the algorithms in terms of permutation error. The lines and shaded regions denote the means and 90\% confidence intervals over 10 runs.}
    \label{fig:natural}
\end{figure}

We apply the NICA algorithms to the problem of recovering images from their mixtures, see Figure~\ref{fig:natural} (left).
Three image patches of size $252\times 252$ pixels were chosen from a set of images of natural scenes \cite{hyvarinen2000emergence} previously used in \cite{hyvarinen2000independent,plumbley2004nonnegative,pehlevan2017blind}. Each image is treated as one source, with the pixel intensities (shifted and scaled to be well-ground and have unit variance) representing the $252^2=63504$ samples. The source vectors were multiplied by a random $3\times 3$ mixing matrix to generate 3-dimensional mixtures, which were presented to the algorithms 5 times with a randomly permuted order in each presentation.
In Figure~\ref{fig:natural} (right), we show the performance of each algorithm.

\section{Summary}
\label{sec:summary}

In this work, we derived 2 algorithms for NICA, each of which can be implemented by biologically plausible single-layer networks. Our networks respectively use two-thirds and one-third as many neurons as the 2-layer biologically plausible network derived in \cite{pehlevan2017blind}.

Our numerical experiments suggest that Algorithms~\ref{alg:online1} and \ref{alg:online2} are outperformed by Nonnegative PCA and the 2-layer NSM. However, a direct comparison is not entirely fair because Nonnegative PCA requires prewhitened inputs and its neural network implementation does not use local learning rules, and the 2-layer NSM network requires 2 layers of neurons. Our algorithms perform both the whitening and the rotation steps in a single layer, which leads to a trade-off in performance. We view this as consistent with the fact that biological systems must make trade-offs between performance and resource efficiency.

Finally, we do not prove convergence guarantees for Algorithms~~\ref{alg:online1} and \ref{alg:online2}. In general, establishing theoretical guarantees for gradient descent-ascent problems is challenging and is further complicated by the non-smoothness of the projected gradient descent steps in Algorithms~\ref{alg:online1} and \ref{alg:online2}.

\paragraph{Acknowledgements.}
We thank Siavash Golkar, Johannes Friedrich, Tiberiu Tesileanu, Alex Genkin, Jason Moore and Yanis Bahroun for helpful comments and feedback on an earlier draft of this work. We especially thank Siavash Golkar for pointing out that, in Sec.~\ref{sec:NICA2}, $\W(\x_t-\ol\x_t)(\x_t-\ol\x_t)$ is not equal to $(\c_t-\ol\c_t)(\x_t-\ol\x_t)$ due to the (suppressed) time-dependency of the weights $\W$.

\bibliographystyle{plain}
\bibliography{blind.bib}

\appendix

\section{Optimization over neural activity matrices $(\Y,\N)$ in the derivation of Algorithm \ref{alg:online1}}
\label{apdx:derivation}

In this section, we show that when $\W_{xy}$ and $\W_{yn}$ are at their optimal values, the optimal neural activity matrices $(\Y,\N)$ can be approximated via projected gradient descent.
We first compute that optimal values of $\W_{xy}$ and $\W_{yn}$.

\begin{lemma}
Suppose $(\W_{xy}^\ast,\W_{yn}^\ast,\Y^\ast,\N^\ast)$ is an optimal solution of objective \eqref{eq:minmax2}.
Then
\begin{align*}
    \W_{xy}^\ast=\P\A^\top,&&\W_{yn}^{\ast,\top}\W_{yn}^\ast=\P\A^\top\A\P^\top,
\end{align*}
for some permutation matrix $\P$.
\end{lemma}

\begin{proof}
From \cite[Theorem 3]{pehlevan2017blind}, we know that every solution of the objective
\begin{align}
    \argmin{\Y\in\R^{d\times T}}-\tr(\delta\Y^\top\delta\Y \delta\X^\top\delta\X)\quad\text{subject to}\quad\delta\Y^\top \delta\Y\preceq T\I_T\quad\text{and}\quad\Y=\F\X,
\end{align}
is of the form $\Y=\F\X$, where $\F$ is a whitening matrix. In particular, since $\Y=\F\A\S$ and $\S$ also has identity covariance matrix, $\Y$ is an orthogonal transformation transformation of $\S$. Furthermore, since $\S$ is well grounded, by \cite[Theorem 1]{plumbley2002conditions}, $\Y$ is nonnegative if and only if $\F\A$ is a permutation matrix. Therefore, every solution $\Y^\ast$ of the objective
\begin{align}
    \argmin{\Y\in\R_+^{d\times T}}-\tr(\delta\Y^\top\delta\Y \delta\X^\top\delta\X)\quad\text{subject to}\quad\delta\Y^\top \delta\Y\preceq T\I_T\quad\text{and}\quad\Y=\F\X,
\end{align}
is of the form $\Y^\ast=\P\X$ for some permutation matrix $\P$.
In addition, differentiating the expression
\begin{align}
    -\tr(\delta\Y^\top\delta\Y \delta\X^\top\delta\X+\delta\N^\top\delta\N(\delta\Y^\top\delta\Y-T\I_T)),
\end{align}
with respect to $\delta\Y$ and setting the derivative equal to zero, we see that the at the optimal value, $\delta\N^{\ast,\top}\delta\N^\ast=\delta\X^\top\delta\X=\delta\S^\top\A^\top\A\delta\S$.

Differentiating $L_1$ with respect to $\W_{xy}$ and $\W_{ny}$, we see that the optimal values for the synaptic weight matrices are achieved at $\W_{xy}=\frac1T\delta\Y\delta\X^\top$ and $\W_{yn}=\frac1T\delta\N\delta\Y^\top$.
Thus,
\begin{align*}
    \W_{xy}^\ast=\frac1T\delta\Y^\ast\delta\X^\top=\frac1T\P\delta\S\delta\S^\top\A^\top=\P\A^\top,
\end{align*}
and
\begin{align*}
    \W_{yn}^{\ast,\top}\W_{yn}^\ast&=\frac{1}{T^2}\delta\Y\delta\N^{\ast,\top}\delta\N^\ast\delta\Y^\top=\frac{1}{T^2}\P\delta\S\delta\S^\top\A^\top\A\delta\S\delta\S^\top\P^\top=\P\A^\top\A\P^\top.
\end{align*}
\end{proof}

Next, we show that when $\W_{xy}$ and $\W_{yn}$ are at their optimal values, the optimal $(\Y^\ast,\N^\ast)$ can be approximated by running the projected gradient dynamics in Eq.~\eqref{eq:Yoffline}.

\begin{lemma}
Suppose $\W_{xy}=\P\A^\top$ and $\W_{yn}^\top\W_{yn}=\P\A^\top\A\P^\top$ for some permutation matrix $\P$.
Then
\begin{align}
    \label{eq:YastNast}
    \Y^\ast=(\W_{yn}^\top\W_{yn})^{-1}\W_{xy}\X=\P\S,&&\N^\ast=\W_{yn}\Y^\ast.
\end{align}
is a solution of the min-max problem
\begin{align}\label{eq:constrained}
    \min_{\Y\in\R_+^{d\times T}}\max_{\N\in\RR^{m
    \times T}}\frac2T\tr\left(\delta\N^\top\W_{yn}\delta\Y-\delta\Y^\top\W_{xy}\delta\X-\delta\N^\top\delta\N\right)\quad\text{s.t.}\quad\Y=\F\X.
\end{align}
In particular, $(\Y^\ast,\N^\ast)$ is the unique solution of the min-max problem
\begin{align}\label{eq:noncentered}
    \min_{\Y\in\R_+^{d\times T}}\max_{\N\in\RR^{m
    \times T}}\frac2T\tr\left(\N^\top\W_{yn}\Y-\Y^\top\W_{xy}\X-\N^\top\N\right),
\end{align}
which can be approximated by running the projected gradient dynamics in Eq.~\eqref{eq:Yoffline}.
\end{lemma}

\begin{proof}
We first relax the condition that $\Y$ be a nonnegative linear transformation of $\X$ and consider the min-max problem
\begin{align*}
    \min_{\Y\in\R^{d\times T}}\max_{\N\in\RR^{d
    \times T}}\frac2T\tr\left(\delta\N^\top\W_{yn}\delta\Y-\delta\Y^\top\W_{xy}\delta\X-\delta\N^\top\delta\N\right).
\end{align*}
After differentiating with respect to $\delta\Y$ and $\delta\N$, we see that this objective is optimized when the centered matrices $\delta\Y$ and $\delta\N$ are given by
\begin{align*}
    \delta\Y=(\W_{yn}^\top\W_{yn})^{-1}\W_{xy}\delta\X,&&\delta\N=\W_{yn}\delta\Y.
\end{align*}
Next, we see that the above relations for the centered matrices hold when $\Y$ and $\N$ are given by Eq.~\eqref{eq:YastNast}, where we have used the fact that $\W_{xy}=\P\A^\top$ and $\W_{yn}^\top\W_{yn}=\P\A^\top\A\P^\top$.
Note that $\Y$ is a linear transformation of $\X$ and $\Y$ is nonnegative since it is a permutation of the nonnegative sources.
It follows that $(\Y,\N)$ is also a solution to the \textit{constrained} min-max problem \eqref{eq:constrained}.
Finally, differentiating the objective in Eq.~\eqref{eq:noncentered} with respect to $\Y$ and $\N$, we see that the optimal $\Y$ and $\N$ are again given by Eq.~\eqref{eq:YastNast}.
\end{proof}

\section{Decoupling the interneuron synapses}
\label{sec:weight}

The NICA algorithm derived in section \ref{sec:NICA1} requires the interneuron-to-output neuron synaptic weight matrix~$\W_{ny}$ to be the the transpose of the output neuron-to-interneuron synaptic weight matrix~$\W_{yn}$. Enforcing this symmetry via a centralized mechanism is not biologically plausible, and is commonly referred to as the weight transport problem.

Here, we show that the symmetry of the 2 weights asymptotically follows from the learning rules in Algorithm \ref{alg:online1}, even when the symmetry does not hold at initialization.
Let $\W_{ny,0}$ and $\W_{yn,0}$ denote the initial values of $\W_{ny}$ and $\W_{yn}$.
Then, in view of the updates rules Algorithm \ref{alg:online2}, the difference $\W_{ny}-\W_{yn}^\top$ after $t$ updates is given by
\begin{equation*}
    \W_{ny}-\W_{yn}^\top = \left(1-\eta\right)^t (\W_{ny,0}-\W_{yn,0}^\top).
\end{equation*}
In particular, the difference decays exponentially.

\section{Details of numerical experiments}
\label{apdx:numerics}

The simulations were performed on an Apple machine with a 2.8 GHz Quad-Core Intel Core i7 processor.

\subsection{Mixing matrices}

We used the $3\times 3$ mixing matrix for the 3-dimensional random uniform sources that was used in \cite{pehlevan2017blind}:
\begin{align*}
    \A=
    \begin{bmatrix}
        0.031518 & 0.38793 & 0.061132 \\
        -0.78502 & 0.16561 & 0.12458 \\
        0.34782 & 0.27295 & 0.67793
    \end{bmatrix}.
\end{align*}
The $10\times 10$ mixing matrix for the 10-dimensional random uniform sources is as follows (entries are rounded to 2 decimal places for space considerations):
\begin{align*}
    \A=
    \begin{bmatrix}
        -1.61 & 0.11  & 0.11  & 1.26  & -0.01 & -1.66 & 0.45  & 0.48  & 0.93  & -0.57 \\
        -0.95 & -0.05 & 0.35  & -0.68 & 1.14  & 0.71  & -0.38 & -0.20 & -0.20 & 2.02  \\
        0.54  & 2.16  & 0.06  & -0.08 & 0.36  & -0.16 & -0.22 & -1.82 & -0.22 & 0.40  \\
        -0.98 & -0.12 & -1.45 & -0.58 & -0.56 & 0.34  & -0.51 & 0.19  & -0.44 & -0.15 \\
        -0.87 & 0.54  & 0.68  & 1.28  & 0.63  & 1.04  & -0.81 & 1.08  & -0.65 & -0.30 \\
        0.91  & 0.84  & 0.45  & -0.31 & -0.14 & -1.46 & -0.18 & 0.48  & -0.41 & 0.75  \\
        -1.20 & 1.29  & 0.39  & -1.40 & 0.84  & -2.32 & -1.54 & -0.26 & -1.99 & -0.34 \\
        1.34  & 0.75  & -1.29 & -0.63 & -1.63 & -1.05 & 0.07  & 0.09  & -0.67 & 0.28  \\
        -0.32 & -0.38 & -0.11 & 1.18  & -0.41 & 0.58  & -0.92 & 1.09  & 0.41  & 1.29  \\
        2.04  & 2.00  & -0.50 & 0.78  & -0.65 & -0.93 & 0.42  & -1.69 & -1.16 & -0.68 
    \end{bmatrix}.
\end{align*}
The $3\times 3$ mixing matrix for the 3-dimensional natural image sources is given by:
\begin{align*}
    \A=
    \begin{bmatrix}
         0.71964649 & -1.55757433 & -1.94561985 \\
        -1.77115767 & -0.99092683 &  0.35559978  \\
        -0.78408667 &  1.09213136 & -1.36539258
    \end{bmatrix}.
\end{align*}

\subsection{Implementation of algorithms}

For each of the algorithms that we implement, we use a time-dependent learning rate of the form:
\begin{align}
    \label{eq:etat}
    \eta_t=\frac{\eta_0}{1+\gamma t}.
\end{align}
To choose the parameters, we perform a grid search over $\eta_0\in\{10^{-1},10^{-2},10^{-3},10^{-4},10^{-5}\}$ and over $\gamma\in\{10^{-2},10^{-3},10^{-4},10^{-5},10^{-6},10^{-7}\}$. In Table \ref{tab:hyper} we report the best performing hyperparameters we found for each algorithm. We now detail our implementation of each algorithm.
\begin{itemize}
    \item [1.] {\bf Bio-NICA with interneurons (Algorithm \ref{alg:online1}):} The neural outputs were computed using the quadratic convex optimization function \texttt{solve\_qp} from the Python package \texttt{quadprog}. After each iteration, we checked if any output neuron had not been active up until that iteration. If so, we flipped the sign of its feedforward inputs. In addition, if the norm of one of the row vectors of $\W_{xy}$ fell below $0.1$, we would replace the row vector with a random vector to avoid the row vector becoming degenerate; and if a singular value of $\W_{xy}$, $\W_{yn}$ or $\W_{ny}$ fell below $0.01$, we replaced the singular value with 1 (we checked every 100 iterations).
    \item [2.] {\bf Bio-NICA with 2-compartmental neurons (Algorithm \ref{alg:online2}):} The neural outputs were computed using the quadratic convex optimization function \texttt{solve\_qp} from the Python package \texttt{quadprog}. We used the time-dependent learning rate of Eq.~\eqref{eq:etat} and included $\tau\in\{0.01,0.03,0.05,0.08,0.1,0.3,0.5,0.8,1,3\}$ in the grid search to find the best performance. After each iteration, we checked if any output neuron had not been active up until that iteration. If so, we flipped the sign of its feedforward inputs. In addition, if a eigenvalue of $\W_{zz}$ fell below $0.01$, we replaced the eigenvalue with 1 to prevent $\W_{zz}$ from becoming degenerate (we checked every 100 iterations).
    \item [3.] {\bf 2-layer NSM:} We implemented the algorithm in \cite{pehlevan2017blind} with time-dependent learning rates. For the whitening layer, we used the optimal time-dependent learning rate reported in \cite{pehlevan2017blind}: $\zeta_t=0.01/(1+0.01t)$. For the NSM layer, we used the time-dependent learning rate of Eq.~\eqref{eq:etat}. To compute the neuronal outputs, we used the quadratic convex optimization function \texttt{solve\_qp} from the Python package \texttt{quadprog}. After each iteration, we checked if any output neuron had not been active up until that iteration. If so, we flipped the sign of its feedforward inputs.
    \item [4.] {\bf Nonnegative PCA (NPCA):} We use the online version given in \cite{plumbley2004nonnegative}.
    The algorithm assumes the inputs are noncentered and whitened. We performed the noncentered whitening offline. After each iteration, we checked if any output neuron had not been active up until that iteration. If so, we flipped the sign of its feedforward inputs.
\end{itemize}

\begin{table}[ht]
    \centering
    \begin{tabular}{|c|c|c|c|c|}
        \hline
                & Alg.~\ref{alg:online1} $(\eta_0,\gamma)$ & Alg.~\ref{alg:online2} $(\eta_0,\gamma,\tau)$ & 2-layer NSM $(\eta_0,\gamma)$ & NPCA $(\eta_0,\gamma)$ \\ \hline\hline
        $d=3$   & $(10^{-2},10^{-3})$                & $(10^{-1},10^{-2},0.8)$     & $(10^{-1},10^{-7})$ & $(10^{-2},10^{-5})$ \\ \hline
        $d=10$  & $(10^{-2},10^{-3})$                & $(10^{-3},10^{-4},0.03)$     & $(10^{-1},10^{-6})$ & $(10^{-2},10^{-5})$ \\ \hline
        Images  & $(10^{-3},10^{-6})$                & $(10^{-2},10^{-4},0.5)$      & $(10^{-1},10^{-6})$  & $(10^{-3},10^{-5})$ \\ \hline
    \end{tabular}
    \vspace{5pt}
    \caption{Optimal hyperparameters used for Alg.~\ref{alg:online1}, Alg.~\ref{alg:online2}, 2-layer NSM, and NPCA.}
    \label{tab:hyper}
\end{table}

\end{document}